\documentclass[journal]{IEEEtran}
\IEEEoverridecommandlockouts
\usepackage{cite}
\usepackage{algorithm,amsbsy,amsmath,amssymb,epsfig,bbm,mathrsfs,multirow,amsthm,amsfonts}
\usepackage{array}
\usepackage{graphicx}
\usepackage{graphics}
\usepackage{subfigure}
\usepackage{epstopdf}
\usepackage{color}
\usepackage{soul}
\usepackage{bbm}

\usepackage{algorithm}
\usepackage[noend]{algpseudocode}

\algrenewcommand\algorithmicforall{\textbf{foreach}}
\algrenewcommand\algorithmicindent{.8em}
\usepackage{mathtools}
\usepackage{pdfpages}

\newcommand{\beq}{\begin{equation}}
\newcommand{\eeq}{\end{equation}}
\newcommand{\bitm}{\begin{itemize}}
\newcommand{\ba}{\begin{array}}
\newcommand{\ea}{\end{array}}
\newcommand{\eitm}{\end{itemize}}
\newcommand{\beqn}{\begin{eqnarray}}
\newcommand{\eeqn}{\end{eqnarray}}
\newcommand{\beqno}{\begin{eqnarray*}}
\newcommand{\eeqno}{\end{eqnarray*}}
\newcommand{\bma}{\begin{displaymath}}
\newcommand{\ema}{\end{displaymath}}
\newcommand{\bnu}{\begin{enumerate}}
\newcommand{\enu}{\end{enumerate}}
\newcommand{\bce}{\begin{center}}
\newcommand{\ece}{\end{center}}
\newcommand{\btb}{\begin{tabular}}
\newcommand{\etb}{\end{tabular}}

\usepackage{graphicx}
\usepackage{textcomp}
\usepackage{xcolor}
\def\BibTeX{{\rm B\kern-.05em{\sc i\kern-.025em b}\kern-.08em
    T\kern-.1667em\lower.7ex\hbox{E}\kern-.125emX}}

\begin{document}

\title{\huge Adversarial Attacks Against Double RIS-Assisted MIMO Systems-based Autoencoder in Finite-Scattering Environments
}

\author{ 
{{Bui Duc Son}, {Ngo Nam Khanh}, Trinh Van Chien,~\IEEEmembership{Member,~IEEE}}, and {Dong In Kim},~\IEEEmembership{Fellow,~IEEE}
\vspace{-1cm}

 \thanks{This research was supported in part by the MSIT (Ministry of Science and ICT), Korea, under the ICT Creative Consilience program (IITP-2020-0-01821) supervised by the IITP (Institute for ICT Planning \& Evaluation). (Corresponding author: Dong In Kim)}  
\thanks{Bui Duc Son and Dong In Kim are with the Department of Electrical and Computer Engineering, Sungkyunkwan University, Suwon 16419, South Korea. Emails: buiducson@skku.edu, dongin@skku.edu.}

\thanks{Ngo Nam Khanh and Trinh Van Chien are with the School of Information and Communication Technology, Hanoi University of Science and Technology, Hanoi 100000, Vietnam. Emails: khanh.nn200318@sis.hust.edu.vn, chientv@soict.hust.edu.vn.}



}

\maketitle
\begin{abstract} 
 Autoencoder permits the end-to-end optimization and design of wireless communication systems to be more beneficial than traditional signal processing. However, this emerging learning-based framework has weaknesses, especially sensitivity to physical attacks. This paper explores adversarial attacks against a double reconfigurable intelligent surface (RIS)-assisted multiple-input and multiple-output (MIMO)-based autoencoder, where an adversary employs encoded and decoded datasets to create adversarial perturbation and fool the system. Because of the complex and dynamic data structures, adversarial attacks are not unique, each having its own benefits. We, therefore, propose three algorithms generating adversarial examples and perturbations to attack the RIS-MIMO-based autoencoder, exploiting the gradient descent and allowing for flexibility via varying the input dimensions. Numerical results show that the proposed adversarial attack-based algorithm significantly degrades the system performance regarding the symbol error rate compared to the jamming attacks.
\end{abstract}

\thispagestyle{empty}
\vspace{-0.5cm}
\section{INTRODUCTION}
\label{sec:introduction}
In the evolution of wireless communication, many techniques have been proposed for better performance regarding throughput, latency, reliability, openness, and security. Multiple-input multiple-output (MIMO) antenna technology exploited in the fifth generation (5G) communication enhances spectral efficiency and reliability thanks to multiplexing and spatial diversity gain. Furthermore, the sixth generation network (6G) is expected to have sub-millisecond latency and support data rates up to one terabit per second with stricter requirements than before \cite{10460991}. Emerging technologies have been anticipated into 6G to boost spectral efficiency and meet these strict communication requirements. Among them, reconfigurable intelligent surface (RIS) has recently demonstrated its potential in enhancing system performance by exploiting passive electronic scattering components to direct electromagnetic waves and achieve a constructive combination at the decoder. 
For harsh propagation environments with weak or even without line-of-sight (LoS), RIS  can guarantee wireless connectivity over extra multi-paths and, therefore, extends transmission coverage under obstacles such as blockage and shadowing \cite{9316920}. Integrating RIS into radio networks and combining it with other advanced wireless variables, including active beamforming vectors, introduce effective resource management strategies. Nonetheless, the optimization problems are often inherently nonconvex and challenging to obtain the global optimum or reach a local optimum with a high implementation cost if the model-based approaches are considered \cite{9316920}. Despite widely investigated in the literature, the model-based approaches are nontrivial to deploy for practical applications due to the fast change of radio channels \cite{9104036}.
\begin{figure}[t]
\centering
\includegraphics[width=1\linewidth]{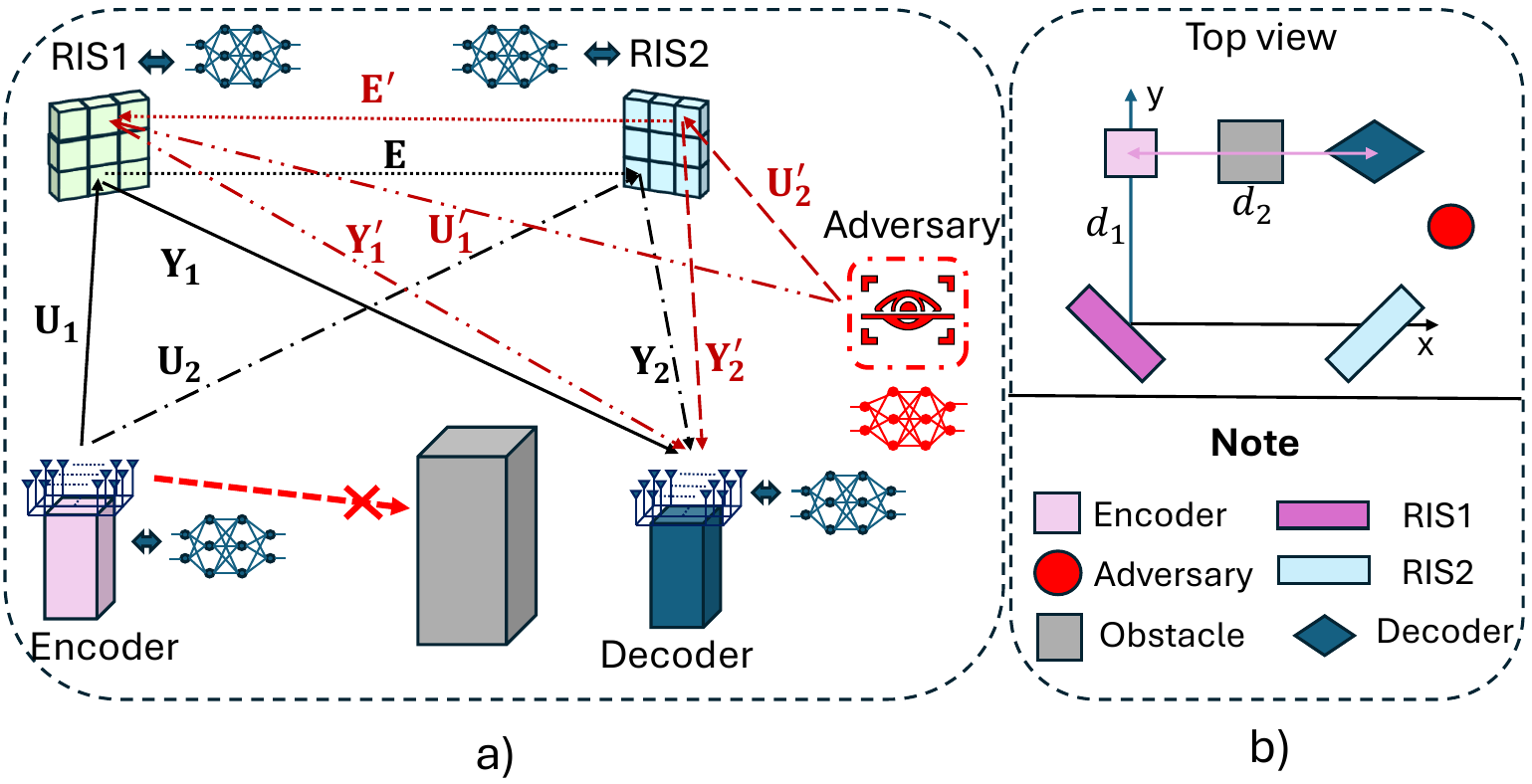}
 \vspace{-0.8cm}
 \caption{(a) The communication model under the adversarial attack. (b) The considered communication model from the top view.}
\label{def}
\vspace{-0.7cm}
\end{figure}

Deep learning, a narrow field of artificial intelligence, has shown its out-performance in enhancing network design and optimizing radio resources \cite{10460991}. 
For example, the authors in \cite{10108002} transformed the active beamforming design of an MU-MISO system into a graph attention network (GAT) to capture the node features affected by inter-link interference. An extension of GAT is to learn the sum-rate optimization problem with the residual-assisted combination and unsupervised learning \cite{10430085}. These graph neural network-based architectures are capable of competing with the model-based approaches under rich scattering environments and without attacks. Notably, autoencoders contribute to the end-to-end learning that obtains the optimal encoding and decoding strategies by a generic artificial intelligent platform instead of specific signal processing modules \cite{9592779}. To minimize the error probability, autoencoders also enhance the trustworthiness and the quality of communication links by involving the channel state information (CSI) into the learning phase and adjusting the transmitted signals accordingly \cite{10136735}. Despite the efficiency and adaptivity, this emerging learning technology is vulnerable to both adversarial and jamming attacks \cite{10460991}. Specifically, the jamming attacks disrupt the communication process, leading to the loss, delay, and corruption of the transmitted signals. Meanwhile, the adversarial attacks fool the autoencoders via adding a small amount of malicious perturbation to the dataset that victims the model with incorrect decisions. The observations  \cite{8651357} indicated that adversarial attacks ruin the end-to-end learning-based communication systems more effectively than jamming attacks. We stress that the previous works have focused on rich scattering environments to demonstrate the strength of the attacking approaches. Nonetheless, the robustness of adversarial attack strategies is under investigation in real propagation environments with limited scatterers.

In this paper,  we investigate the abilities of adversarial attacks against a double RIS-MIMO system-based autoencoder over spatially correlated fading channels with a limited number of scatterers utilizing a white box scenario. Our main contributions are summarized as follows: $i)$ We introduce an autoencoder-based model with legitimate learning-based devices that can replace the RIS-aided MIMO system and suffer from attackers. Our considered system operates with the presence of finite scatterers; $ii)$ We propose a white-box adversarial attack to reduce communication reliability. The attacking algorithm can effectively learn and damage the autoencoder based on the projected gradient descent (PGD) and prior channel state information; and $iii)$ Numerical results show that the proposed adversarial attack algorithm can increase the error probability of an RIS-aided MIMO system more effectively than the state-of-the-art benchmarks.
\vspace{-0.4cm}

\section{SYSTEM MODEL}
The considered RIS-assisted MIMO system is illustrated in Fig.~\ref{def} comprising the encoder, RIS~$1$, RIS~$2$, decoder, and adversary. RIS~$1$ and RIS~$2$ are, respectively, placed near the encoder and decoder to achieve the cooperative passive beamforming gain instead of a single RIS \cite{zheng2021double}. We employ one-dimensional convolutional neural networks (CNNs) to replace the physical components of these blocks, creating an autoencoder system.\footnote{The autoencoder is trained in the unsupervised learning fashion based on the transmit data and channel state information to predict the system parameters including the phase shift coefficients. With an intelligent reconfiguration, the phase shift coefficients are predicted by a generic CNN circuit and sent to the RISs. This implementation keeps the RISs passive and steps toward practical applications.} Aligned with natural propagation environments, the channels follow spatially correlated fading with a finite number of scatterers.
\subsection{Double scattering fading channels}
 In this subsection, we describe the practical fading model, which has finite scatterers on the transmitter and receiver \cite{ 9531522}. 
 Let us denote $\mathbf{U}_{i} \in \mathbb{C}^{A_i \times N_{t}}$, $\mathbf{Y}_i \in \mathbb{C}^{ N_{r} \times A_i}, i = 1, 2$, and $\mathbf{E} \in \mathbb{C}^{A_2 \times A_1}$  are the channels between the encoder and the RIS~$i$, the decoder and the RIS~$i$, and between  the two RISs, respectively. Here, $N_{t}$ and $N_{r}$ are the number of antennas equipped at the encoder and decoder; $A_i$ represents the number of passive reflection elements of the RIS $i$. Assuming that the channels appear to be flat in the frequency domain and static in the time domain \cite{9531522}, which are denoted as $\mathbf{N} \in \{ \mathbf{U}_i, \mathbf{Y}_i, \mathbf{E} \}$, $ i = 1, 2$ and modelled as
\begin{equation}
    \mathbf{N} = \sqrt{\Omega}\left( \sqrt{\frac{\chi}{\chi +1}}\overline{\mathbf{N}} + \sqrt{\frac{1}{\chi +1}}\hat{\mathbf{N}}\right),
\end{equation}
where ${\mathbf{\overline{N}}}$ and ${\mathbf{\hat{N}}}$ are the  LoS and NLoS components; $\Omega$ and $\chi$ denote the distance-dependent and LoS-dominant factors. We assume that the antennas of the encoder and decoder are arranged in a uniform linear array (ULA), while the passive reflecting elements of RISs are organized in a uniform planar array (UPA). Consequently, the LoS channels are given as
\begin{align}
& \overline{\mathbf{U}}_{i} = \mathbf{a}_{i}(\gamma^A_{\rm 1Ti},\theta^A_{\rm Ti})(\mathbf{a_{U}}_{i}(\gamma^D_{\rm Ti}))^T \in \mathbb{C}^{A_{i} \times N_{t}},\\
& \overline{\mathbf{Y}}_{i} = \mathbf{a}_{{\mathbf{Y}}_{i}}(\gamma^A_{\rm Ri})(\mathbf{a}_{i}(\gamma^D_{\rm Ri},\theta^D_{\rm Ri}))^T \in \mathbb{C}^{N_{r} \times A_{i}},\\
&\overline{\mathbf{E}} = \mathbf{a}_{2}(\gamma^A_{\rm R},\theta^A_{\rm R}) (\mathbf{a}_{1}(\gamma^D_{\rm Ti},\theta^D_{\rm 1}))^T \in \mathbb{C}^{A_{2} \times A_{1}},
\end{align}
where $\gamma$ and $\theta$ are the azimuth and elevation angles; The superscripts $A$ and $D$ are angle-of-arrival and angle-of-departure;  $\mathbf{a}_{\mathbf{U}_i} \in \mathbb{C}^{N_t \times 1}$ and $\mathbf{a}_{\mathbf{Y}_i} \in \mathbb{C}^{N_t \times 1}$ are the ULA response vectors of the encoder and decoder; and $\mathbf{a}_i$ is the UPA response vector of RIS~$i$ with $\mathbf{a}_i = \mathbf{a}_{vi} \otimes \mathbf{a}_{hi}$, where $\otimes$ denotes the Kronecker product and $\mathbf{a}_{vi}$ and $\mathbf{a}_{hi}$ are the array response vectors along the two axes with lengths $A_{vi}$ and $A_{hi}$, respectively. Here, \(A_{i} = A_{vi} \times A_{hi}\) represents the number of elements in each row and column in the UPA arrangement of the respective RIS. The NLoS channels $  \widehat{\mathbf{N}} \in \{ \mathbf{\hat{U}}_{i}, \mathbf{\hat{Y}}_{i}, \mathbf{\hat{E}} \}$ with \( \widehat{\mathbf{N}} \in \mathbb{C}^{N_1 \times N_2} \) and \( N_1, N_2 \in \{N_{t}, N_{r}, A_1, A_2\} \), can be defined as 
$\widehat{\mathbf{N}} = \mathsf{SC}_{\mathbf{N}}^{-0.5}\mathbf{R}^{0.5}_{t,\mathbf{N}}\mathbf{Q}_{\mathbf{N}}\mathbf{S}^{0.5}\mathbf{P}_{\mathbf{N}}\mathbf{R}^{0.5}_{r,\mathbf{N}}$,
where the subscript $\mathbf{N} \in \{ \mathbf{U}_i, \mathbf{Y}_i, \mathbf{E} \}$ denotes the specific channel link;  $\mathsf{SC}_{\mathbf{N}}$ represents the number of scatterers associated with $\mathbf{N}$; $\mathbf{R}_{r,\mathbf{N}} \in \mathbb{C}^{N_2 \times N_2}$, $\mathbf{S}_{\mathbf{N}} \in \mathbb{C}^{\mathsf{SC}_{\mathbf{N}} \times \mathsf{SC}_{\mathbf{N}}}$, and $\mathbf{R}_{t,\mathbf{N}} \in \mathbb{C}^{N_1 \times N_1}$ are  the decoder, scatterer, and encoder correlation matrices for $\mathbf{N}$; $\mathbf{Q_N} \in \mathbb{C}^{N_1 \times \mathsf{SC}_{\mathbf{N}}}$ and $\mathbf{P_N} \in \mathbb{C}^{\mathsf{SC}_{\mathbf{N}} \times N_2}$ represent the small-scale fading between the transmitter and receiver sides and their corresponding scattering clusters. Specifically, the correlation matrices between the encoder and scatterers are 
\vspace{-0.2cm}
\begin{equation}
    \label{CorMatrix}
    [\mathbf{R}_{t,\rm \mathbf{U}_{i}}]_{m,n}=\mathsf{SC}_{\mathbf{ U}_{i}^{-1}}  \sum\nolimits_{k=-b}^{b} \exp \left(j2 \pi a_{\mathrm s} q \sin(a_\mathrm v) \right),
\end{equation}
where \([\mathbf{R}_{t, \mathbf{U}_i}]_{m,n}\) indicates the \((m,n)\)-th element of the matrix \(\mathbf{R}_{t,\mathbf{U}_i}\); \(b = 0.5(\mathsf{SC}_{\mathbf{U}_i} - 1)\) and \(q = m - n\); \(\mathsf{SC}_{\mathbf{U}_i}\) denotes the number of scatterings corresponding to channel \(\mathbf{U}_i\); \(a_\mathrm{s}\) is the antenna spacing at the transmitter side; \(a_{\mathrm v} = \frac{k\Psi_e}{1 - \mathbf{SC}_{\mathbf{U}_i}}\), where \(\Psi_e\) is the angular spread of the signals. Similar to \eqref{CorMatrix}, the correlation matrices between the decoder and the scatterer can be calculated with the corresponding parameters. The correlation matrices along each axis of the RIS are given as
\begin{align}
   & [\mathbf{R}_{\mathrm{v}}^i]_{m,n}=\mathsf{SC}_{\mathbf{N}}^{-1}  \sum\nolimits_{k=-a}^{a} \exp \left(j2 \pi d_{\mathrm{v}} q \sin \beta \right), \\
   & [\mathbf{R}_{\mathrm{h}}^i]_{m,n}=\mathsf{SC}_{\mathbf{N}}^{-1}  \sum\nolimits_{k=-a}^{a} \exp \left(j2 \pi d_{\mathrm{h}} q \sin \beta \right),
\end{align}
where $a =0.5(\mathsf{SC}_{\mathbf{N}} - 1)$ and $\beta = k \Psi_r(1 - \mathsf{SC}_{\mathbf{N}})^{-1}$; $d_\mathrm{v}$ and $d_\mathrm{h}$ indicate the distances along the vertical and horizontal directions between two adjacent reflecting elements of RIS~$i$. The correlation matrices among the scattering points are
\begin{equation}
    [ \mathbf{S_{\mathbf{N}}}]_{m,n}=\mathsf{SC}_{\mathbf{N}}^{-1} \sum\nolimits_{k=-a}^{a} \exp \left(j2 \pi d_\mathrm{sc} q \sin \beta' \right),
\end{equation}
where $\beta' =  k \Psi_\mathrm{sc}(1 - \mathsf{SC}_{\mathbf{N}}^{-1})$; $d_\mathrm{sc}$ and $\Psi_\mathrm{sc}$ are the distance between two scatterers and the angular spread.
\vspace{-0.6cm}
\subsection{Encoder}
The encoder is designed using a one-dimensional convolutional neural network (1D-CNN) to substitute all the traditional components. The input of the encoder is a bit string $\mathbf{i}_k$, represented by a one-hot vector of length $M$.\footnote{The one-hot vector is a vector of length $M$ corresponding to the $M$  modulated signals in the constellation. One-hot means this vector's $(M-1)/M$ elements equal zero, whereas only the remaining equals one.} This one corresponds to a modulated signal of a constellation, for example, $M$-QAM (quadrature amplitude modulation). The input is a sequence denoted by $ \mathbf{I}_d = [\mathbf{i}_1, \ldots, \mathbf{i}_{B_L}] \in \mathbb{C}^{M \times B_L}$, where $B_L$ represents the 1D-CNN's block length. After being fed into the neural network, $\mathbf{I}_d$ is processed by multiple one-dimension convolution (Conv1D) layers with the rectified linear unit (ReLU) and the batch normalization (BN). Before transmission, the encoded signals are normalized by a layer called power normalization, a custom layer with non-trainable parameters. The output of the encoder, denoted as $\mathbf{O}_E = P\mathbf{O}_E^{'}\big(\sqrt{\mathbb{E}[|\mathbf{O}_E^{'}|]^2}\big)^{-1} \in \mathbb{C}^{N_{t} \times B_L}$ will be transmitted by $N_{t}$ antennas, where $\mathbf{O}_E^{'} \in \mathbb{C}^{N_{t} \times B_L}$ is the output of the last 1D-CNN layer and $P$ represents the  transmitted power.
\vspace{-0.6cm}
\subsection{RIS~1 and RIS~2}
We exploit 1D-CNN to learn the double-RIS network behaviors. The input of the RIS~1 and the RIS~2 are denoted as $\mathbf{R_1} = [\mathbf{r}_{11}, \ldots, \mathbf{r}_{1  {B_L}}] \in \mathbb{C}^{A_{1} \times B_L}$ and $\mathbf{R}_2 = [\mathbf{r}_{21}, \ldots, \mathbf{r}_{2  {B_L}}] \in \mathbb{C}^{A_{2} \times B_L}$, respectively. For the RIS~1, the elements of matrix $\mathbf{R}_1$ are $\mathbf{r}_{1i} = \mathbf{{U}^i_1}\mathbf{o}_i$, where $i = 1, \ldots, B_L$  and  $\mathbf{o}_i$ are the channels and the output of the encoder at the $i$-th symbol. A matrix of size $2A_1 \times B_L$ is constructed by separating and reshaping the real and image components of the input signals. It is then processed through several 1D-CNN layers to obtain the predicted phase shift vector of the RIS~1, denoted as $\pmb{\Gamma}_1 = [\pmb{\gamma}_{11}, \ldots, \pmb{\gamma}_{1{B_L}}] \in \mathbb{C}^{A_{1} \times B_L}$. Here, $\pmb{\gamma}_{1i} = \{ \gamma_{11}^i, \ldots, \gamma_{1 A_{1}}^i \} $ is utilized to define the predicted reflection matrix $\pmb{\psi}_1^i = \mathrm{diag} (e^{j\gamma_{11}^i}, \ldots, e^{j\gamma_{1{A_1}}^i} )$. For the RIS~2, the elements of matrix $\mathbf{R}_2$ are defined as $\mathbf{r}_{2i} = (\mathbf{U}^i_2 + \mathbf{E}^i \pmb{\psi}_1^i \mathbf{U}^i_1)\mathbf{o}_i$ with $i = 1, 2, \ldots B_L$. Then, the predicted phase shift vector of the RIS~2 is $\pmb{\Gamma}_2 = [\pmb{\gamma}_{21}, \ldots, \pmb{\gamma}_{2  {B_L}}] \in \mathbb{C}^{A_{2} \times B_L}$. The reflection matrix at the $i$-th symbol of the RIS~2 is  $\pmb{\psi}_2^i = \mathrm{diag} (e^{j\gamma_{21}^i}, \ldots, e^{j\gamma_{2{A_2}}^i} )$, where $\pmb{\gamma}_{2i} = \{ \gamma_{21}^i, \ldots, \gamma_{2 {A_2}}^i \}$.
\vspace{-0.6cm}
\subsection{Decoder}
 The predicted phase-shift vectors and propagation channels are involved in the received signal as follows:
\vspace{-0.2cm}
\begin{equation}
    \begin{aligned}
    \mathbf{r}_i & = (\mathbf{Y_2}^i \pmb{\psi}_2^i \mathbf{E}^i \pmb{\psi}_1^i \mathbf{U}^i_1 + \mathbf{Y_1}^i \pmb{\psi}_1^i \mathbf{U}^i_1 + \mathbf{Y_2}^i \pmb{\psi}_2^i \mathbf{U}^i_2)\mathbf{o}_i \\
                &= \mathbf{K^i}\mathbf{o}_i, \quad i = 1, \ldots, B_L.
    \end{aligned}
    \vspace{-0.2cm}
\end{equation}
When the attack is not performed after receiving $B_L$ symbols, $ \mathbf{S} = \{ \mathbf{S}^1, \ldots, \mathbf{S}^{B_L} \}\in \mathbb{C}^{N_{t} \times N_{r}} \times B_L$ and $\mathbf{R} = [ \mathbf{r}_1, \ldots, \mathbf{r}_{B_L}]$ represent the cascaded channel and the received signal, respectively. These parameters are combined to construct a matrix with the shape $(N_{r} + N_{r}N_{t}) \times B_L$ to be fed into the decoder. The real and image components of the input are manipulated to create the matrix of size $2(N_{r} + N_{r}N_{t}) \times B_L$, which is then processed through several 1D-CNN layers followed by the BN and ReLU layers. Finally, a softmax layer \cite{iwana2019explaining} is employed to convert the output to the matrix $\mathbf{T} = [\mathbf{t}_1, \ldots, \mathbf{t}_{B_L}] \in \mathbb{C}^{M \times B_L}$, where each $\mathbf{t_i}$ presents the vector containing all probabilities of possible messages at the $i$-th symbol. The decoded message $\mathbf{\hat{I}_D} = [\mathbf{\hat{i}}_1, \ldots, \mathbf{\hat{i}}_{B_L}]$ is determined based on the index of $\mathbf{t_i}$, which is the maximum probability.

\vspace{-0.4cm}
\subsection{Adversary}
\small
\begin{algorithm}[t]
\caption{RIS-MIMO adversarial example-based PGD (RMAEP)}
\label{alg:2}
\begin{algorithmic}[1]
\State \% \textit{Main Algorithm}
\State \textbf{Input:} Model architecture, ${p}_{\rm{PSR}}$, $\sigma^2$,${\{\mathbf{G}^{i}\}}$.
\State {Set} $n_p$ and $\mathbf{p}_{\rm{adv}}= {\mathbf{0}}$.
\ForAll{$i$ in range($n_p$)}
    \State Choose $\mathbf{I}_r \in \mathcal{I} $ uniformly at random; 
    \State ${\tilde{\mathbf{p}}_{\rm{adv}}= \mathbf{G}^{i} \mathbf{p}_{\rm{adv}}}$;
    \State Set a random noise ${\mathbf{n}} \sim \mathcal{CN}(0, \sigma^2\mathbf{I})$;
    \State Set $\hat{\mathbf{I}}_r = \mathcal{D}(\mathcal{R}_2(\mathcal{R}_1(\mathcal{E}(\mathbf{I}_r))) + \mathbf{n} + {\tilde{\mathbf{p}}_{\rm{adv}}}) = D(\mathbf{w}_{\rm adv})$ ;
    \If{$\hat{\mathbf{I}}_r = \mathbf{I}_r$}
        \State $\mathbf{w_{\rm adv}} \leftarrow \mathbf{w_{\rm adv}} +  {\tilde{\mathbf{p}}_{\rm{adv}}}$;
        \State $\tilde{\mathbf{p}}_{\mathrm{add}} \leftarrow$ \textit{PGD\_based\_Method($\mathbf{w_{\rm adv}}$)};
        \State {$\mathbf{p}_{\mathrm{add}} = ((\mathbf{G}^{i})^H \mathbf{G}^i)^{-1}  (\mathbf{G}^i)^H  \tilde{\mathbf{p}}_{\mathrm{add}}$};
        \If{$\|\mathbf{p}_{\mathrm{adv}} + \mathbf{p}_{\mathrm{add}}\|_2^2 \leq {p}_{\mathrm{PSR}}$}
            \State $\mathbf{p}_{\mathrm{adv}} \leftarrow$  $\mathbf{p}_{\mathrm{adv}} + \mathbf{p}_{\mathrm{add}}$;
        \Else
            \State $\mathbf{p}_{\mathrm{adv}} \leftarrow \sqrt{p_{\mathrm{PSR}}} (\mathbf{p}_{\mathrm{adv}} + \mathbf{p}_{\mathrm{add}})/\|\mathbf{p}_{\mathrm{adv}} + \mathbf{p}_{\mathrm{add}}\|_2$;
        \EndIf
    \EndIf
\EndFor
\State \textbf{Output:} Adversarial perturbation vector $\mathbf{p}_{\mathrm{adv}}.$
\\

\% \textit{Sub Algorithm: PGD\_based\_Method for Problem~\ref{constrain}}
\State \textbf{Input: } {$\mathbf{w} =  \mathcal{R}_2(\mathcal{R}_1(\mathcal{E}(\mathbf{I}_r))) + \mathbf{n}.$}
\State {Init: }$\epsilon \leftarrow \mathbf{0}^{{M} \times B_L}$;
\ForAll{$i$ in range(${{M}}$)}
    \State $\epsilon_{\min} \leftarrow 0$ and $\epsilon_{\max} \leftarrow p_{\max}$;
    \State $\mathbf{p}_{\mathrm{norm}} \leftarrow {\nabla_\mathbf{w} \mathcal{L}(\mathbf{w}, e_{i})}/{\|\nabla_\mathbf{w} \mathcal{L}(\mathbf{w}, e_{i})\|_2}$;
    \While{$\epsilon_{\max} - \epsilon_{\min} > \epsilon_{\mathrm{acc}}$}
        \State $\mathbf{p}_{\mathrm{temp}} \leftarrow \mathbf{p}_{\mathrm{norm}}$ and $\mathbf{w_{\rm adv}} \leftarrow \mathbf{w}$;
        \ForAll{$j$ in range($n_s$)}
            \State $\epsilon_{\mathrm{ave}} \leftarrow (\epsilon_{\max} - \epsilon_{\min})/2$;
            \State $\mathbf{w_{\rm adv}} \leftarrow \mathbf{w_{\rm adv}} - (\epsilon_{\mathrm{ave}}/n_s) \mathbf{p}_{\mathrm{temp}}$; 
            \State$\mathbf{w_{\rm adv}} \leftarrow \pmb{\Pi_{\beta }}(\mathbf{w_{\rm adv}})$;
            \State $\mathbf{p}_{\mathrm{temp}} \leftarrow \nabla_{\mathbf{w_{\rm adv}}} \mathcal{L}(\mathbf{w_{\rm adv}}, e_{i})/\|\nabla_{\mathbf{w_{\rm adv}}} \mathcal{L}(\mathbf{w_{\rm adv}}, e_{i})\|_2$;
        \EndFor
        \State $\mathbf{p}_{\mathrm{norm}} \leftarrow \mathbf{p}_{\mathrm{temp}}$;
        \If{$D(\mathbf{w_{\rm adv}}) == y_{\mathrm{label}}$}
            \State $\epsilon_{\max} \leftarrow \epsilon_{\mathrm{ave}}$;
        \Else
            \State $\epsilon_{\min} \leftarrow \epsilon_{\mathrm{ave}}$;
        \EndIf   
    \EndWhile
    \State $\epsilon_{i}$ $\leftarrow$ $\epsilon_{\max}$;
\EndFor
\State $\mathrm{target} \leftarrow \arg \min \epsilon$ and $\epsilon^* \leftarrow \min \epsilon$;
\State $\mathbf{p}_{\mathrm{add}} \leftarrow {\epsilon^*\nabla_\mathbf{w} \mathcal{L}(\mathbf{w}, e_{\text{target}})}/{\|\nabla_\mathbf{w} \mathcal{L}(\mathbf{w}, e_{\mathrm{target}})\|_2}$;
\State \textbf{Output:} Additional adversarial perturbation vector $\mathbf{p}_{\mathrm{add}}$.
\end{algorithmic}
\end{algorithm}
\normalsize
{By smart signal processing, the attacker can ruin the received data as}
\begin{equation}
        {{\mathbf{\hat{r}}_i} = {\mathbf{{r}}_i} + \mathbf{G}^i \mathbf{p}_{\mathrm{adv}},}
\end{equation} 
{where $\mathbf{p}_{\mathrm{adv}}$ is the adversarial perturbation vector; $\mathbf{G}^i= \mathbf{Y'_1}^i \pmb{\psi}_1^i \mathbf{E'}^i \pmb{\psi}_2^i \mathbf{U'}^i_2 + \mathbf{Y'_1}^i \pmb{\psi}_1^i \mathbf{U'}^i_1 + \mathbf{Y'_2}^i \pmb{\psi}_2^i \mathbf{U'}^i_2$ is the aggregated channel between the adversary and the decoder with the presence of both RISs, in which $\mathbf{U'}_{i} \in \mathbb{C}^{A_i \times N_{t}}$, $\mathbf{Y'}_i \in \mathbb{C}^{ N_{r} \times A_i}, i = 1, 2$, and $\mathbf{E'} \in \mathbb{C}^{A_1 \times A_2}$  are the attack channels between the adversary and the RIS~$i$, the decoder and the RIS~$i$, and between the two RISs, respectively, and follow double scattering fading channel.} After collecting ${B}_L$ received symbols, the perturbed received signals $\mathbf{\hat{R}} = [ \mathbf{\hat{r}}_1, \ldots, \mathbf{\hat{r}}_{B_L}]$ and $\mathbf{S}$ are connected to form the input data of the encoder resulting in a matrix with the shape ${(N_{r} + N_{r}N_{t})} \times B_L$. The real and image parts of the input are utilized to generate a matrix of size $2{(N_{r} + N_{r}N_{t})} \times B_L$. 
This matrix then undergoes a sequence of 1D-CNN layers, followed by the BN and ReLU layers. Finally, the softmax layer converts the output to the matrix $\hat{\mathbf{T}} = [\mathbf{\hat{t}}_1, \ldots, \mathbf{\hat{t}}_{B_L}] \in \mathbb{C}^{{M} \times B_L}$, where $\mathbf{t}_i$ represents the vector containing all possible message probabilities at the $i$-th symbol. The decoded message $\mathbf{\hat{I'}}_D = [\hat{\mathbf{i}}'_1, \ldots, \hat{\mathbf{i}}'_{B_L}]$ is determined based on the index of $\mathbf{\hat{t}_i}$ with the maximum probability. {In this paper, we will study the effectiveness of adversarial attacks in damaging legitimate devices, including the support of RISs by exploiting deep learning. We stress that a powerful adversarial attack makes the system less reliable, consequently causing a high symbol error rate (SER).}
\vspace{-0.2cm}
\section{Adversarial attack}

We now propose an algorithm that an adversary can exploit to degrade the autoencoders with the white-box attack.
\begin{figure*}[t]
	\centering
    \begin{minipage}[t]{0.24\textwidth}
	\includegraphics[width=1\textwidth]{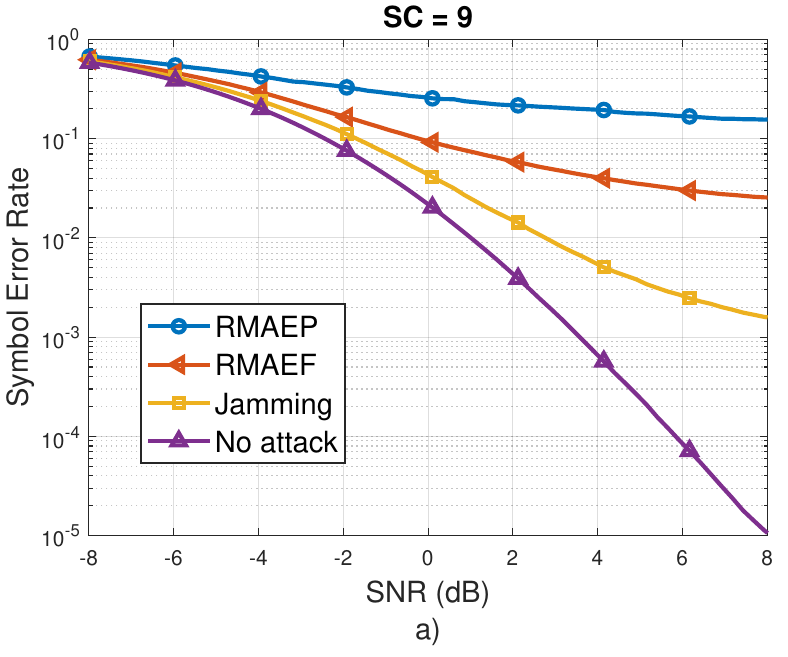}
    \end{minipage}
    \begin{minipage}[t]{0.24\textwidth}
	\includegraphics[width=1\textwidth]{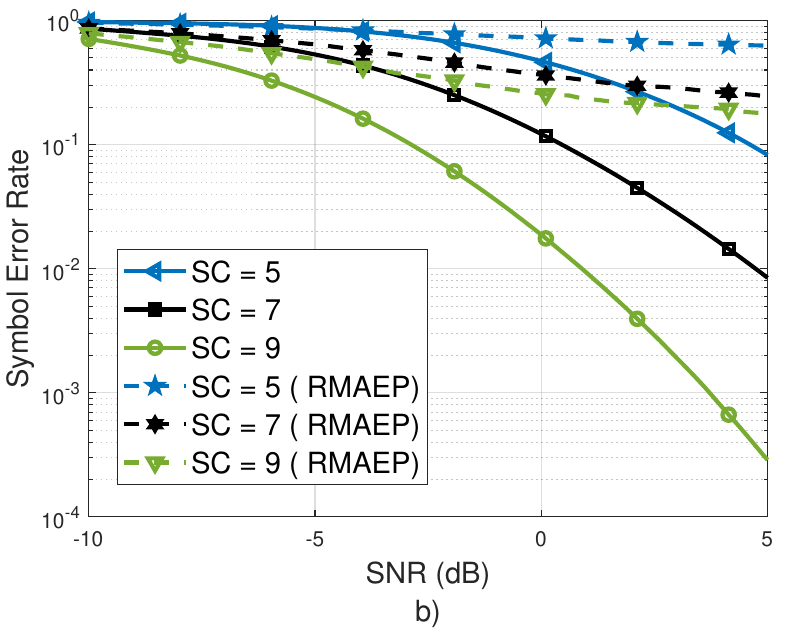}
    \end{minipage}
    \begin{minipage}[t]{0.24\textwidth}
	\includegraphics[width=1\textwidth]{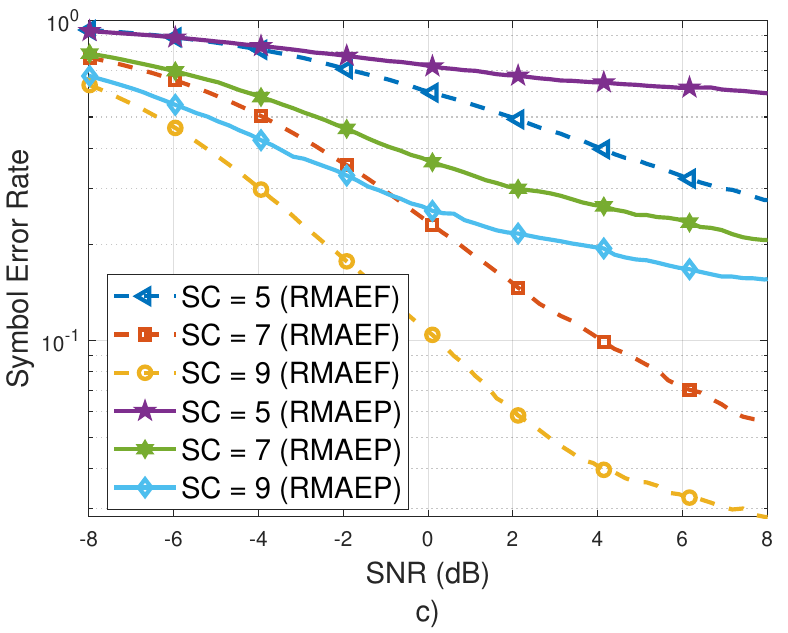}
    \end{minipage}
    \begin{minipage}[t]{0.24\textwidth}
	\includegraphics[width=1\textwidth]{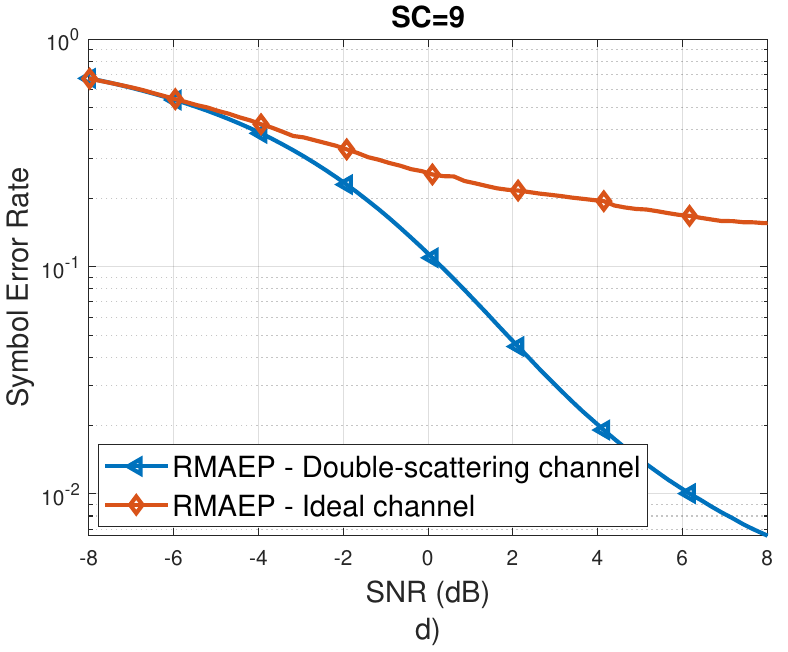}
    \end{minipage}
\vspace{-0.4cm}
\caption{(a) The SER of the four considered benchmarks when the number of scatterers is 9 {under the ideal attacking channel}. (b) The SER of RMAEP with the different number of scatterers {under the ideal attacking channel}. (c) Comparing the SER between RMAEP and RMAEF with the different numbers of scatterers {under the ideal attacking channel}. {(d) Comparing the SER of RMAEP under ideal attacking and double-scattering channels.}}
\vspace{-0.5cm}
\label{convergence1}
\end{figure*}
Based on the network architecture in Fig.~\ref{def}(a), an adversary uses perturbations to fool the decoder. Mathematically modelling this approach, we define
\begin{align}
&\mathbf{w} =  \mathcal{R}_2(\mathcal{R}_1(\mathcal{E}(\mathbf{I}_r))) + \mathbf{n}, \label{eq:w1}\\
&\mathbf{w}_{\rm adv} = \mathcal{R}_2(\mathcal{R}_1(\mathcal{E}(\mathbf{I}_r))) + \mathbf{n} +  {\tilde{\mathbf{p}}_{\rm{adv}}}, \label{eq:w2}
\end{align}
as the secured and perturbed signals observed at the decoder, respectively. In \eqref{eq:w1} and \eqref{eq:w2},  $\mathcal{R}_1(\cdot)$, $\mathcal{R}_2(\cdot)$, and $\mathcal{E}(\cdot)$ represent the autoencoder through which the signals are processed and conditioned on the channels. For each signal $\mathbf{w}$,  ${\tilde{\mathbf{p}}_{\rm{adv}}}$  in \eqref{eq:w2} is obtained by solving the following optimization problem
\begin{equation}
\label{constrain}
    \begin{aligned}
         & \underset{\mathbf{p}_{\rm adv}}{\mathrm{minimize}} \quad \| \mathbf{p}_{\rm adv} \|_2 \\ 
         & \text{subject to} \quad \mathcal{D}(\mathbf{w}) \neq \mathcal{D}({\mathbf{w}_{\rm adv})},
    \end{aligned}
\end{equation}
where $\mathcal{D}(\cdot)$ is the decoder$; {\tilde{\mathbf{p}}_{\rm{adv}}= \mathbf{G}^{i} \mathbf{p}_{\rm{adv}}}$. We emphasize that the adversary cannot directly utilize the solution of problem~\eqref{constrain} as an optimal adversarial perturbation to poison the dataset because it depends on a specific input setting. Although we examine white-box scenarios in this paper, the attacker cannot know exactly what and when data will be transmitted. For this reason, we propose Algorithm~\ref{alg:2}, named RIS-MIMO adversarial example-based PGD (RMAEP), to optimize the adversarial perturbation vector $(\mathbf{p}_{\rm adv})$ that damages dataset in the RIS-MIMO system. The algorithm begins with initializing the $\mathbf{p}_{\rm adv}$ and the number of iterations $n_p$. We also denote the dataset $\mathcal{I} \triangleq \{\mathbf{I}_d\}$ contains $Z$ input matrices of the encoder. In each iteration, one randomly selects a data sample $\mathbf{I}_r $ from $ \mathcal{I} = [\mathbf{I_1}, \ldots, \mathbf{I_{Z}}] \in \mathbb{C}^{{M} \times B_L \times 
Z} $, and concentrates it with random noise $\mathbf{n} \sim \mathcal{CN}(0, \sigma^2 \mathbf{I}) $. Thus, the model output, \(\hat{\mathbf{I}}_r\), is formulated as 
$\hat{\mathbf{I}}_r = \mathcal{D}(\mathcal{R}_2(\mathcal{R}_1(\mathcal{E}(\mathbf{I}_r))) + \mathbf{n} +  {\tilde{\mathbf{p}}_{\rm{adv}}})$. 
If the model is not affected by the perturbation, i.e., \(\hat{\mathbf{I}}_r = \mathbf{I}_r\), $\mathbf{p}_{\rm adv}$ will be updated by adding a small amount of perturbation denoted as \(\mathbf{p}_{\mathrm{add}}\). The value of \(\mathbf{p}_{\mathrm{add}}\) as well as an approximate solution for \eqref{constrain} is computed by exploiting the PGD. Then, \(\pmb{\epsilon} \) is initialized as a null matrix of size \({M} \times L \), matching the input data. Next, the proposed algorithm repeats for a specified number of iterations, denoted by \(e_i\), being assigned a random label from the dataset. In each iteration, it determines the minimum and maximum values for \(\epsilon_i\) and uses a binary search combined with the PGD to adjust \(\epsilon_i\) towards the change in prediction compared to the desired incorrect label. In each individual step of the binary search to determine the appropriate value, we include an additional loop to find the optimal \(w_{\rm adv}\). In this process, \(\pmb{\Pi}_{\pmb{\beta}}\) is defined as
\begin{align}
\pmb{\Pi}_{\pmb{\beta}} =
\begin{cases}
    \mathbf{w}_{\mathrm{adv}} \leftarrow \mathbf{w} - \pmb{\beta}, & \text{if } \|\mathbf{w}_{\mathrm{adv}}\|_2 < \|\pmb{\alpha}_{\mathrm{low}}\|_2, \\
    \mathbf{w}_{\mathrm{adv}} \leftarrow \mathbf{w}_{\mathrm{adv}}, & \text{if } \|\pmb{\alpha}_{\mathrm{low}}\|_2 \leq \| \mathbf{w}_{\mathrm{adv}}\|_2 \leq \|\pmb{\alpha}_{\mathrm{up}}\|_2, \\
    \mathbf{w}_{\mathrm{adv}} \leftarrow \mathbf{w} + \pmb{\beta}, & \text{if } \|\mathbf{w}_{\mathrm{adv}}\|_2 > \|\pmb{\alpha}_{\mathrm{up}} \|_2,
\end{cases}
\label{eq:labelone}
\end{align}
where $\mathbf{w} - \pmb{\beta} = \pmb{\alpha}_{\mathrm{low}}$ and $\mathbf{w} + \pmb{\beta} = \pmb{\alpha}_{\mathrm{up}}$; $\epsilon_{\mathrm{ave}} = (\epsilon_{\max} - \epsilon_{\min})/2$; $\pmb{\beta} = \frac{\epsilon_\mathrm{ave}}{n_s} \mathbf{p}_{\mathrm{norm}}$; $n_s$ is the number of optimization steps. After determining the optimum for each \(\epsilon_i\), the proposed algorithm selects \(\epsilon^\ast\) as the smallest value and evaluates additional adversarial perturbation vector \(\mathbf{p}_{\mathrm{add}}\) based on the gradient of the loss function {($\mathcal{L}$)} at the target point, which equals to $ \arg \min \epsilon$, please refer to line 35 to 36 of Algorithm~\ref{alg:2} for more details. The ultimate solution, \(\mathbf{p}_{\mathrm{add}}\), is considered as additive noise to the input data. Note that $\mathbf{p}_{\mathrm{adv}} \leftarrow \mathbf{p}_{\mathrm{adv}} + \mathbf{p}_{\mathrm{add}}$ is updated based on checking the $\ell_2$-norm of the total noise vector. The procedure is repeated until all the \(n_p\) iterations are completed and the final adversarial perturbation vector \(\mathbf{p}_{\mathrm{adv}}\) is returned. Using this iterative process, we can generate a vector  \( \mathbf{p}_{\mathrm{adv}} \), which can universally attack the input of the autoencoder. Alternatively, the developed perturbation can be applied to multiple different inputs without needing to be adjusted for each case.  
As illustrated in the sub-algorithm of Algorithm~\ref{alg:2},  a proper $\mathbf{p}_{\mathrm{norm}}$ is found based on the PGD, enabling it to generate a more effective adversarial perturbation vector iteratively.
The computational complexity of RMAEP is in the order of 
\(\mathcal{O}(N_p N_s M^2_QB^2_L)\).
 {Previous works have studied the effectiveness of adversarial attacks in wireless communications, for example \cite{8651357}. Nonetheless, the authors in \cite{8651357} only focused on adversarial attacks for a single-input and single-output system under additive white Gaussian noise and an ideal attacking channel gain. In contrast, we investigate the abilities of adversarial attacks for RIS-aided MIMO systems under practical conditions, including spatial correlation and a limited number of scatterers. More complicated than \cite{8651357},  the autoencoder enables active and passive beamforming designs for fading channels. The considered systems introduce new optimization variables, e.g., phase shifts, and new challenges, e.g., propagation channels following uncommon distributions.}

\vspace{-0.5cm}

\section{Numerical results}
As shown in Fig.~\ref{def}(b), the coordinates of the encoder, the decoder, and the RISs are $(0, d_1, d_H)$, $(d_2, d_1, 0)$, $(0, 0, d_H)$, and $(d_2, 0, 0)$, respectively. Here, $d_1$ is the distance between the encoder and the RIS~1;  $d_2$ is the distance from the RIS~1 to the RIS~2; and $d_H$ is the height of the encoder, the RIS~1, and the RIS~2 compared to the decoder as in reality. We set $d_1 = 100$~[m], $ d_2 = 200$~[m], and  $d_H = 2$~[m] for Monte Carlo simulations. For antenna arrays and phase shifts, we select $N_{t} = N_{r} = 16$, $A_{1} = A_{2} = 32$, ${M} = 64$, and $B_L = 20$. During the training phase, 100,000 symbol data samples from $\mathcal{I}$ are exploited to train the autoencoder. Besides, 10,000 random data samples are created for the testing process. The SNR for training is 15~[dB], and the LoS-dominant factor is 0.2. For attacking, we establish ${p}_{\rm{PSR}}$ as -7dB. The number of random samples taken, $n_p$, is 50, whereas the number of iterations for the adversarial sample generation process Algorithm~\ref{alg:2}, $n_s$, is $20$. All layers in the 1D-CNNs utilize Adam optimization with an initial learning rate of $0.001$. The number of training epochs is $1000$, and {$\mathcal{L}$} is binary cross entropy. Monte-Carlo simulations are conducted on a computer featuring a 12th Generation Intel(R) Core(TM) i7-12700 processor running at 2.10 GHz and an NVIDIA GeForce RTX 4090 Ti boasting 24GB of memory. The effectiveness of RMAEP is demonstrated over the other benchmarks comprising crafting physical adversarial perturbations, denoted as RMAEF \cite{8651357},  the jamming attack \cite{9779086}, and the secured system (no attack) \cite{10136735}. 

In Fig.~\ref{convergence1}(a), all the attackers can increase the SER of the autoencoder system compared to a secured system. As the SNR grows, the SER of the secured system declines sharply from nearly $10^0$ at $-8$~[dB] to $10^{-5}$ at $8$~[dB]. Whereas the jamming attack shows notable performance, increasing the system's SER to $3 \times 10^{-3}$ at $8$~[dB], the RMAEF and RMAEP show even more significant effects, raising the metric to around $4 \times 10^{-2}$ and approximately $2 \times 10^{-1}$ at the same SNR. Next, Fig~\ref{convergence1}(b) displays the ability of the RMAEP to damage the autoencoder with different numbers of scatterers. In general, the SER reduces as the number of scatterers increases \cite{10136735}. Aligned with real environments with a small number of scatterers, the SER is high at the low SNRs but declines sharply as the strength of the received signal improves. When the autoencoder is attacked by RMAEP, the SER is always high and much worse than a secured system. The numerical results manifest the effectiveness of Algorithm~\ref{alg:2} in finite scattering environments. Fig.~\ref{convergence1}(c) shows the SER of the two adversarial attackers. Even though the performance of RMAEF is notable, it is less effective than RMAEP. {Finally, Fig.~\ref{convergence1}(d) compares the performance of RMAEP under ideal attacking and double-scattering channels. Although the SER of RMAEP under the double-scattering channel is worse than the ideal attacking channel, it is still better than the jamming attack under the ideal attacking channel. Specifically, at 6 [dB], the SER of RMAEP under the double-scattering channel and jamming attack under the ideal attacking channel are $5 \times 10^{-3}$ and $10^{-2}$, respectively.}
\vspace{-0.2cm}
\section{Conclusion}
\vspace{-0.2cm}
This paper investigated adversarial attacks against the double RIS-assisted MIMO systems-based autoencoders in scattering environments with LoS and NLoS components. Despite the complicated structure of finite scatterers with multiple randomnesses,  we proposed three algorithms to craft adversarial perturbation to the victim models. The SER was exploited as a measurement metric to evaluate the system performance of different attacking benchmarks. Numerical results showed that our proposed adversarial attack algorithms outperform the traditional jamming attack in the white box scenarios by effectively decreasing communication reliability. {Besides, some countermeasures could be used for future work against the proposed attack, such as adversarial training, defensive distillation, and adversarial denoising \cite{10460991}. The RIS-empowered physical layer security scheme-based secrecy rate optimization\cite{10143983} is also promising.}
\vspace{-0.0cm}





\bibliographystyle{IEEEtran}
\bibliography{IEEE}
\end{document}